\documentclass[%
reprint,
superscriptaddress,
 amsmath,amssymb,
 aps,
 pra,
]{revtex4-1}

\usepackage{float}
\usepackage{graphicx}
\usepackage{dcolumn}
\usepackage{bm}

\usepackage[utf8]{inputenc} 

\usepackage{multirow}
\usepackage[warn]{mathtext}
\usepackage{amsfonts}
\usepackage{amsmath}
\usepackage{amssymb}
\usepackage{braket}
\usepackage{bbold}
\usepackage{textcomp} 
\usepackage{indentfirst} 
\usepackage{amsmath} 
\usepackage{graphicx}

\DeclareGraphicsExtensions{.pdf,.png,.jpg}

\usepackage{hyperref}

\usepackage{algpseudocode}

\usepackage[nottoc]{tocbibind}

\usepackage{xcolor}
\hypersetup{
    colorlinks,
    linkcolor={red!50!black},
    citecolor={blue!50!black},
    urlcolor={blue!80!black}
}



\begin{document}

    \title{States dressing analysis in a transmon-transmon-bus system}

    \author{P.A.~Gladilovich}
    \email{pa.gladilovich@misis.ru}
    \affiliation{National University of Science and Technology MISIS, Moscow 119049, Russia}
 
    \author{R.A.~Migdisov}
    \affiliation{National University of Science and Technology MISIS, Moscow 119049, Russia}

    \author{A.V.~Sabluk}
    \affiliation{National University of Science and Technology MISIS, Moscow 119049, Russia}
    
    \author{P.S.~Burtsev}
     \affiliation{National University of Science and Technology MISIS, Moscow 119049, Russia}

    \author{N.N.~Abramov}
    \affiliation{National University of Science and Technology MISIS, Moscow 119049, Russia}

    \author{N.Y.~Rudenko}
    \affiliation{National University of Science and Technology MISIS, Moscow 119049, Russia}

    \author{V.I.~Chichkov}
    \affiliation{National University of Science and Technology MISIS, Moscow 119049, Russia}

    \author{N.A.~Maleeva}
    \affiliation{National University of Science and Technology MISIS, Moscow 119049, Russia}

\date{\today}
 
	\begin{abstract}
	The multi-qubit gates fidelity of superconducting quantum processors can be limited due to the dressing of computational states by noncomputational ones. Here, we experimentally and analytically investigate a transmon-transmon-bus system where the computational states dressing is tunable over a broad range. We estimate the dressing using three methods: a full three-element model, an effective mode approach, and an unperturbed mode approach. The obtained results highlight the importance of the accurate estimation and control of the computational states dressing in order to optimize  gates on superconducting platform.
    \end{abstract}
	
\maketitle

\section{Introduction} 
Superconducting qubits are now one of the leading platforms for building scalable quantum processors. Their success comes from a favorable combination of long coherence times, flexible control, mature fabrication technology, and other advantages \cite{Lang2025,Mayer2025,Bland2025,Koch2007,Arute2019,Kjaergaard2020}. High gate fidelity, especially for two-qubit gates, remains a key requirement on the path toward universal quantum computing \cite{Arute2019, Acharya2024}.

Most modern two-qubit gates use an additional coupling element between qubits. The coupler can be static \cite{Majer2007, Kandala2019, Besedin2026, Xiong2026, Song2025} or tunable \cite{Ding2023, Li2024, Lin2025, Zhang2024, Moskalenko2022, Renger2026}. Furthermore, the coupling mechanism can involve a single \cite{Moskalenko2022} or multiple \cite{Egorova2025} modes of the coupler. In all these cases, the coupler states are non-computational and effect the computational states of the coupled qubits.

Typically, this influence manifests as: (I)~leakage into the noncomputational subspace \cite{Yang2024, Berezkin2025, Marxer2025} (a dynamic process); and (II)~a degradation of the coherent properties of the dressed \cite{Cohen1992, Wilson2007} (i.e., static dressing) states within the computational subspace \cite{Marxer2025, Marxer2023}. Both processes limit the fidelity of the executed gates.

Advanced leakage-suppression protocols, such as Phase-Averaged Leakage Error Amplification (PALEA), can reduce the contribution of parasitic dynamics to the two-qubit gate error budget down to 15\%. At the same time, insufficient control over the bare-state composition of dressed computational states can increase the incoherent-error contribution to the total error budget up to 30\% \cite{Marxer2025}.

The contribution of non-computational states to dressed computational states is not always directly accessible in experiment. It reflects the spatial delocalization of the dressed-state wave function over the physical elements of the system \cite{Jingjing2023}. In this work, we study three numerical methods for estimating the dressing \(D\) of computational states by a single non-computational state. We use a three-element superconducting circuit comprising a transmon qubit, a transmon coupler, and a $\lambda / 2$ resonator. Frequency-domain measurements show excellent agreement between experiment and the model for both mode-frequency dynamics and effective couplings. Using this model, we compare the exact value of \(D\) in the full system with two simplified approaches based on effective and unperturbed computational modes. Crucially, we find that the critical dressing coincides with an inversion of the spectral ordering and the equality of the effective couplings. Our results show that accurate estimation and control of computational subspace dressing are essential for optimizing logical operations in superconducting quantum processors.

\section{Problem Statement and System Architecture}

The system consists of three elements coupled in series: a qubit (Q), a coupler (C), and a half-wavelength resonator (bus). A dressed state \(|\psi^{(k)}\rangle_{\text{dressed}}\) can be expanded in the basis of bare states, i.e., the basis of noninteracting elements:
\begin{eqnarray}\label{eq0}
 \ket{\psi^{(k)}} _{\text{dressed}} = \sum_{\mathcal{Q, C, B}} c_{\mathcal{Q, C, B}}^{(k)} |\mathcal{Q, C, B}\rangle_{\text{bare}},\nonumber\\ 
\quad \sum_{\mathcal{Q, C, B}} |c_{\mathcal{Q, C, B}}^{(k)}|^2 = 1.
\end{eqnarray}
Here, the coefficients \(c_{\mathcal{Q, C, B}}^{(k)}\) give the contribution of the bare states \(\ket{\mathcal{Q,C,B}}_{\text{bare}}\) to the dressed state. Below, we omit the subscript \textit{bare} and write \(|\mathcal{Q, C, B}\rangle_{\text{bare}}=|\mathcal{Q, C, B}\rangle\). If only the first excited states of Q, C, and bus are considered, Eq.~\eqref{eq0} reduces to
\begin{eqnarray}\label{eq0.25} 
|\psi^{(k)}\rangle_{\text{dressed}} = c_{e, g, 0}^{(k)} |e, g, 0\rangle + \nonumber\\
c_{g, e, 0}^{(k)} |g, e, 0\rangle + c_{g, g, 1}^{(k)} |g, g, 1\rangle, \nonumber\\
\quad |c_{e, g, 0}^{(k)}|^2 + |c_{g, e, 0}^{(k)}|^2 + |c_{g, g, 1}^{(k)}|^2 = 1.
\end{eqnarray}
We name dressed states according to their dominant basis vector. For example, for the bus state,
\begin{eqnarray}\label{eq0.5} 
|bus\rangle = c_{e, g, 0} |e, g, 0\rangle +\nonumber \\
c_{g, e, 0} |g, e, 0\rangle + c_{g, g, 1} |g, g, 1\rangle,\nonumber \\
\quad |c_{e, g, 0}|^2 + |c_{g, e, 0}|^2 < |c_{g, g, 1}|^2.
\end{eqnarray}
This notation requires inequalities analogous to Eq.~\eqref{eq0.5}.

In our device, Q and C are implemented as Xmons \cite{Barends_2013} with two-junction asymmetric SQUIDs. The bus is a section of a coplanar transmission line, as shown in Fig.~1. The states of Q and C are read out through individual quarter-wavelength resonators coupled to a common readout line. Flux biasing and microwave driving of Q and C are performed through individual control lines.

\begin{equation}
\label{eq1}
\begin{aligned}
\widehat{H} ={} &
\sum_{i \in \{Q,C\}} \Big[
4E_{C}^{(i)}\widehat{n}_{(i)}^{2} \\[4pt]
&\quad - \big(E_{J}^{(i)} + E_{j}^{(i)}\big)
\cos \widehat{\varphi}_{(i)}
\cos\!\left(\pi \frac{\Phi_{(i)}}{\Phi_0}\right) \\[4pt]
&\quad + \big(E_{J}^{(i)} - E_{j}^{(i)}\big)
\sin \widehat{\varphi}_{(i)}
\sin\!\left(\pi \frac{\Phi_{(i)}}{\Phi_0}\right)
\Big] \\[4pt]
& + \hbar \omega_{\mathrm{bus}}\, \widehat{a}^{\dagger}\widehat{a}
+ g_{Q\text{-}C}\, \widehat{n}_{Q}\widehat{n}_{C} \\
& + g_{C\text{-}\mathrm{bus}}\, \widehat{n}_{C}\widehat{n}_{\mathrm{bus}}
+ g_{Q\text{-}\mathrm{bus}}\, \widehat{n}_{Q}\widehat{n}_{\mathrm{bus}}.
\end{aligned}
\end{equation}

The full Hamiltonian in Eq.~\eqref{eq1} includes the energies of the individual subsystems and the couplings between them. The fundamental mode of the half-wavelength coplanar resonator (bus) is modeled as an LC oscillator with frequency \(\omega_{\text{bus}}/2\pi = 4.238\) GHz. The capacitances to ground \(C_{Q-\text{gnd}}\), \(C_{C-\text{gnd}}\), and the SQUID junction parameters set the charging energies \(E_{C}^{Q}=E_{C}^{C}=285\) MHz and the Josephson energies \(E_{J}^{Q}=4.7\) GHz, \(E_{j}^{Q}=9.1\) GHz, \(E_{J}^{C}=4.4\) GHz, \(E_{j}^{C}=8.8\) GHz. These parameters determine the flux-tunable frequency ranges \(f_Q \in [2.821, 5.307]\) GHz and \(f_C \in [2.834, 5.185]\) GHz. The mutual capacitances \(C_{ Q-C }\), \(C_{C-\text{bus}}\), and \(C_{Q-\text{bus}}\) define the direct capacitive couplings \(g_{Q-C}=7.861\) MHz, \(g_{C-\text{bus}}=14.675\) MHz, and \(g_{Q-\text{bus}}=0.234\) MHz. All numerical values were extracted from spectroscopic measurements of the sample shown in Fig.~1(b).

\begin{figure}
\centering
\includegraphics[width=0.48\textwidth]{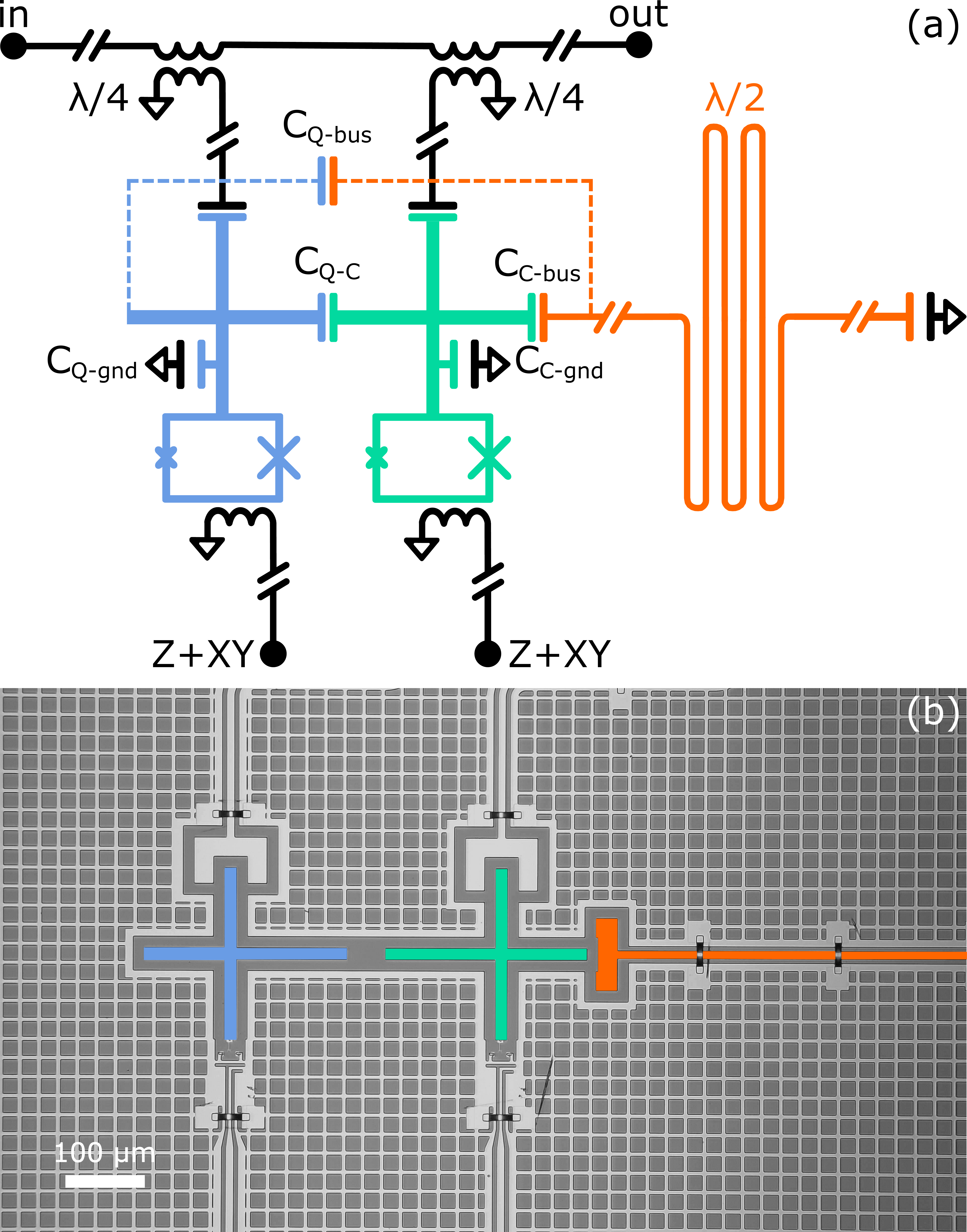}
\caption{\label{fig:1}(a) Circuit diagram and (b) false-colored optical micrograph of the system. The qubit (Q) is shown in blue, the coupler (C) in green, and the half-wavelength resonator (bus) in orange. The individual readout \(\lambda/4\) resonators are capacitively coupled to Q and C and are shown in black. Both resonators are coupled to a coplanar in-out transmission line. The Z+XY lines for flux biasing and microwave driving are inductively coupled to the SQUIDs of Q and C.}
\end{figure}

\begin{figure}
\centering
\includegraphics[width=0.48\textwidth]{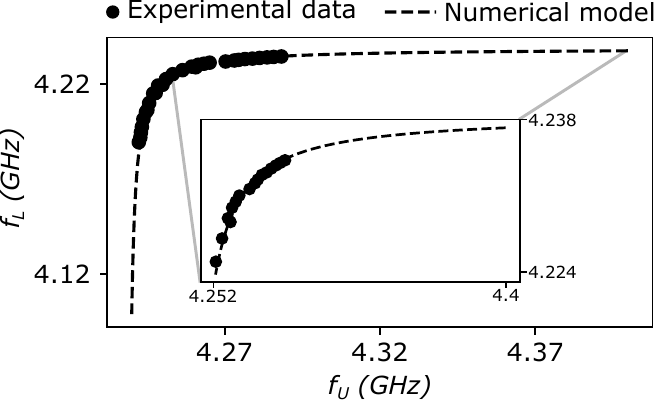}
\caption{\label{fig:2} Experimental data (points) and model data (dashed line) for the lower-mode frequency (L) as a function of the upper-mode frequency (U) of the C-bus subsystem under flux biasing of C. The inset shows the frequency region up to the critical value \(f_U = f_U^c\).}
\end{figure}

\begin{figure*}
\centering
\includegraphics[width=0.98\textwidth]{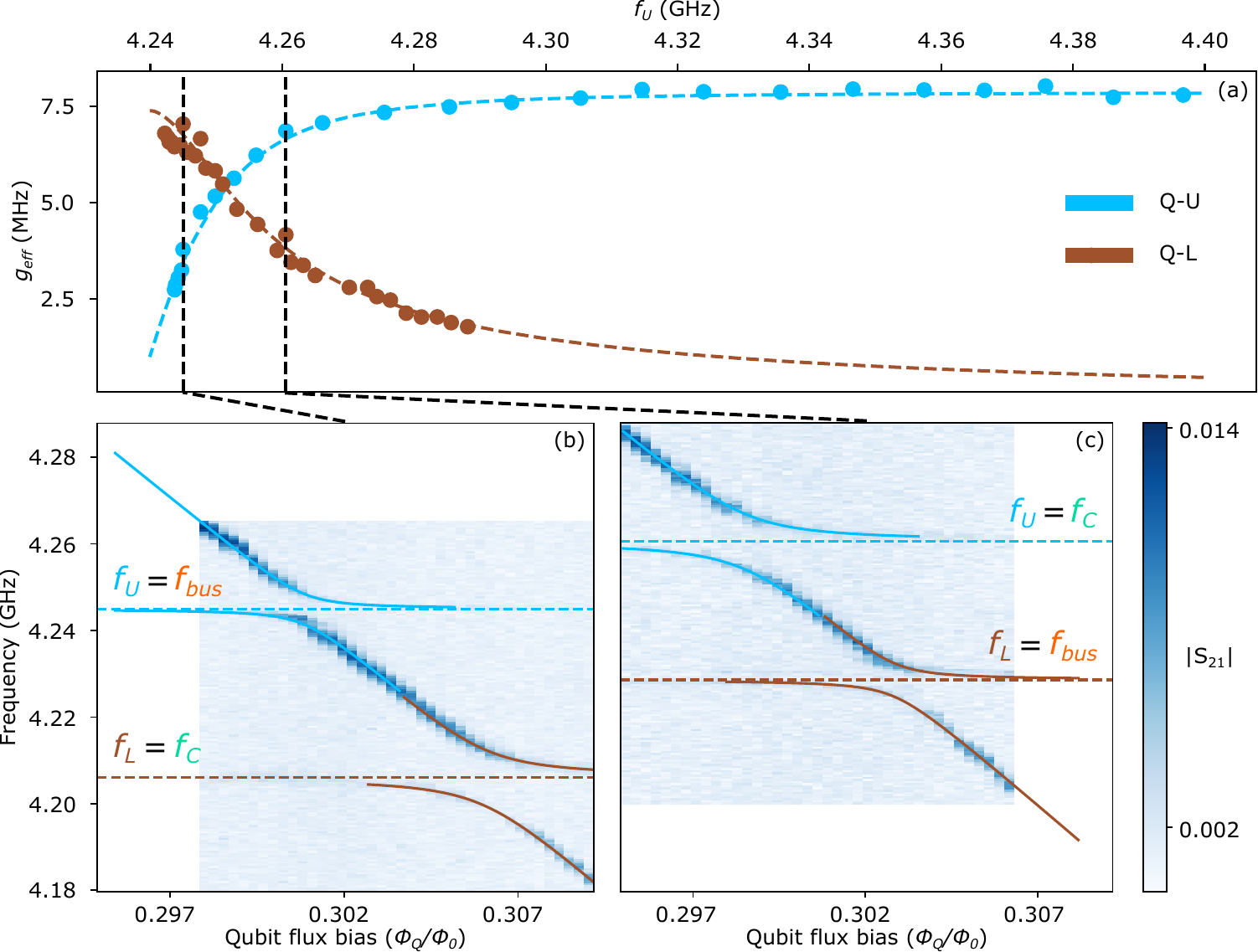}
\caption{\label{fig:3} Comparison of experimental and model data for the effective coupling of qubit Q to the upper (U) and lower (L) modes of the C-bus subsystem. (a) Effective couplings \(g_{\text{eff}}^{Q-U}\) (sky blue) and \(g_{\text{eff}}^{Q-L}\) (brown) as functions of the upper-mode frequency \(f_U\). Dashed lines show model data from Eq.~(4), and points show experimental data. (b), (c) Examples of experimental and model spectra after (b) and before (c) inversion of the spectral order.}
\end{figure*}

\begin{figure*}
\centering
\includegraphics[width=0.98\textwidth]{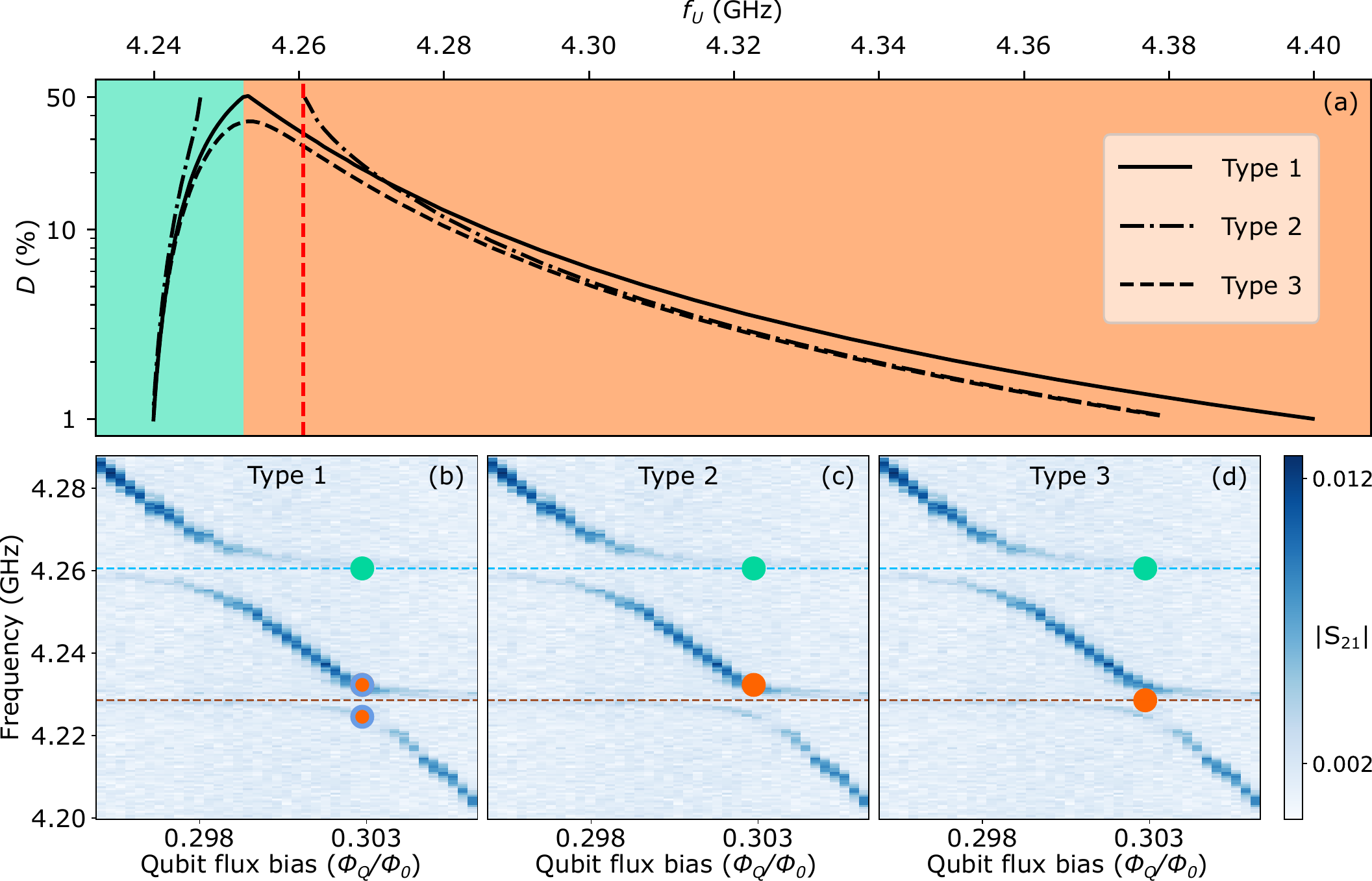}
\caption{\label{fig:4} Comparison of three methods for estimating the dressing \(D\) of resonant computational states \(|Q\rangle\) and \(|L\rangle\). (a) \(D\) as a function of the upper-mode frequency \(f_U\), calculated in the full system (type 1, solid line), with an effective mode (type 2, dash-dotted line), and with an unperturbed mode (type 3, dashed line). The red dashed line marks the value of \(f_U\) used to compare the three methods. The background colors show the regions where the dressed computational state \(|L\rangle\) corresponds to the dressed state \(|bus\rangle\) (orange) or \(|C\rangle\) (green). (b--d) Spectra with schematic illustrations of the three methods: (b) the exact full-system method, (c) the effective-mode method, where the computational mode is taken as the mode of the Q-bus avoided crossing closest to C, and (d) the unperturbed-mode method, where the computational mode is taken at the center of the Q-bus avoided crossing.}
\end{figure*}

We verified that Hamiltonian \eqref{eq1} accurately describes the system in Fig.~1 using two experiments. In both experiments, Q served as a scanning element near the upper (U) and lower (L) modes of the C-bus subsystem. Obtained avoided crossings allowed us to determine the mode positions (Fig.~2) and their effective coupling \(g_{\text{eff}}\) (Fig.~3) at different flux biases applied to the SQUID of C. Like any dressed states, \(|U\rangle\) and \(|L\rangle\) can be expanded in the bare-state basis \(|e, g, 0\rangle\), \(|g, e, 0\rangle\), and \(|g, g, 1\rangle\), as in Eq.~\eqref{eq0.5}. As the flux bias of C changes, the mode frequencies change as well. The bare-state composition of \(|U\rangle\) and \(|L\rangle\), determined by \(c_{e, g, 0}\), \(c_{g, e, 0}\), and \(c_{g, g, 1}\), therefore also changes. Section~2 gives a detailed analysis of this composition. Here we only note the initial configuration: C lies above the bus in frequency, so \(|U\rangle \leftrightarrow |C\rangle \approx|g, e, 0\rangle\) and \(|L\rangle \leftrightarrow |bus\rangle \approx|g, g, 1\rangle\).

In the first experiment, Q scanned the lower mode of the C-bus subsystem, denoted L. Initially, this mode corresponds to the bus. The scan was repeated at different flux biases of coupler C. The avoided crossing between Q and L gives both the mode frequency \(f_L\) and the effective coupling \(g_{\text{eff}}^{Q-L}\), as shown in Fig.~3(a). At the same time, we tracked the frequency \(f_U\) of the upper mode, which initially corresponds to C. The points in Fig.~2 show measured dependence of \(f_L\) on \(f_U\).

In the second experiment, Q scanned the upper mode of the C-bus subsystem, denoted U. Initially, this mode corresponds to C. By varying the external flux applied to C, we extracted \(f_U\) from the position of the Q-U avoided crossing and determined the effective coupling \(g_{\text{eff}}^{Q-U}\), as shown in Fig.~3(a). We could not track mode L over the full bias range of C because the bus has no individual readout resonator (starting from a certain bias point, mode L corresponds to the bus).

Figure~3(a) summarizes the experimental results. It shows \(g_{\text{eff}}^{Q-L}\) and \(g_{\text{eff}}^{Q-U}\) as functions of the upper-mode frequency \(f_U\). The data reveal two distinct frequency regions. In the high-frequency region, \(f_U > f_U^c \approx 4.252\) GHz, \(g_{\text{eff}}^{Q-U}\) is larger than \(g_{\text{eff}}^{Q-L}\). At the upper edge of this region, near \(f_U \approx 4.4\) GHz, we find \(g_{\text{eff}}^{Q-U} \approx g_{Q-C}\) and \(g_{\text{eff}}^{Q-L} \approx g_{Q-\text{bus}}\). This confirms the initial frequency ordering shown in Fig.~3(c). The inset of Fig.~2 shows the corresponding mode-frequency dynamics for \(f_U > f_U^c\). The lower-mode frequency changes from \(f_L\approx4.238\) GHz at the upper edge of the range to \(f_L\approx4.224\) GHz at \(f_U=f_U^c\). This shift is exactly what one expects from hybridization of the bus mode from \(f_{\text{bus}}=\omega_{\text{bus}}/2\pi\) to \(f_{\text{bus}}=\omega_{\text{bus}}/2\pi-g_{C-\text{bus}}\) as C approaches the bus from above in frequency. The critical point is \(f_{C}^{c}=\omega_{\text{bus}}/2\pi+g_{C-\text{bus}}\approx 4.252\) GHz.

In the low-frequency region, \(f_U < f_U^c\), the effective-coupling hierarchy is inverted: \(g_{\text{eff}}^{Q-U} < g_{\text{eff}}^{Q-L}\). At the lower edge of this region, the system reaches \(g_{\text{eff}}^{Q-L} \approx g_{Q-C}\) and \(g_{\text{eff}}^{Q-U} \approx g_{Q-\text{bus}}\). This indicates inversion of the initial frequency ordering of C and bus, as shown in Fig.~3(b). At \(f_U^c\), the two effective couplings are equal: \(g_{\text{eff}}^{Q-L}=g_{\text{eff}}^{Q-U}=g_{\text{eff}}^c \approx 5.2\) MHz. Across the full frequency range, the experimental data agree very well with numerical simulations based on Eq.~\eqref{eq1}. The agreement holds for both mode-frequency dynamics (Fig.~2) and effective couplings (Fig.~3). This validates the model and allows us to use it to analyze the bare-state composition of the dressed states.

\section{Methods for Estimating Non-Computational-State Dressing of Computational States}

For quantum computing, it is important to quantify how strongly unwanted, non-computational states participate in the dynamics of the working states. In this work, following Eq.~\eqref{eq0.25}, the quantity \(D\) denotes the participation weight \(|c_{\mathcal{Q, C, B}}|^2\) of the dominant basis vector \(|\mathcal{Q,C,B}\rangle\) of the dressed non-computational state \(|U\rangle\) in the resonant computational states \(|Q\rangle\) and \(|L\rangle\), as shown in Fig.~4(a). This spectral configuration is relevant for two-qubit operations such as iSWAP \cite{Moskalenko2022}. In such gates, the computational modes are tuned into resonance to maximize their interaction, while a nearby non-computational mode provides the required coupling and dresses the working pair. Thus, \(D\) is equal to either \(|c_{g, e, 0}|^2\) or \(|c_{g, g, 1}|^2\) in the expansions of \(|Q\rangle\) and \(|L\rangle\). The choice depends on the dominant basis vector of the dressed state \(|U\rangle\), namely \(|g,e,0\rangle\) or \(|g,g,1\rangle\), respectively.

We use \(D=50\%\) as a conventional threshold for a change in the dominant basis vector of a dressed state. Above this value, a state identified with one element, for example \(|bus\rangle\) in Eq.~\eqref{eq0.5}, becomes closer in composition to the state of another element. For instance, it becomes closer to \(|C\rangle\) when \(|c_{e, g, 0}|^2 + |c_{g, g, 1}|^2 < |c_{g, e, 0}|^2\).

We demonstrate the three methods using the data shown in Fig.~3(c). At this frequency point, all three methods correctly identify the correspondence \(|U\rangle \leftrightarrow |C\rangle\) and \(|L\rangle \leftrightarrow |bus\rangle\). The parameters are \(f_U=f_C = 4.261\) GHz, \(f_L=f_{\text{bus}} = 4.229\) GHz, \(g_{\text{eff}}^{Q-U}=g_{\text{eff}}^{Q-C} = 6.867\) MHz, and \(g_{\text{eff}}^{Q-L}=g_{\text{eff}}^{Q-\text{bus}} = 4.172\) MHz. This point is marked by the red dashed line in Fig.~4(a).

The first method is the full-system calculation, illustrated in Fig.~4(b). It treats the complete three-element system, with Q tuned into resonance with the bus. Coupler C lies above them in frequency and produces the effective coupling between Q and bus. This coupling creates the frequency splitting \(2g_{\text{eff}}^{Q-\text{bus}}\). In this regime, Hamiltonian \eqref{eq1} reduces to
\begin{equation}\label{eq2} 
\hat{H}_{\text{full}}=h\begin{pmatrix}
f_Q^{'} & g_{\text{eff}}^{Q-C} & g_{\text{eff}}^{Q-\text{bus}}\\
g_{\text{eff}}^{Q-C} & f_C^{'} & g_{\text{eff}}^{C-\text{bus}}\\
g_{\text{eff}}^{Q-\text{bus}} & g_{\text{eff}}^{C-\text{bus}} & f_{\text{bus}}^{'}
\end{pmatrix}.
\end{equation}

Solving the eigenvalue problem gives the dressed eigenstates \(|C\rangle\), \(|Q\rangle\), and \(|bus\rangle\), with frequencies \(f_C\), \(f_{\text{bus}}+g_{\text{eff}}^{Q-\text{bus}}\), and \(f_{\text{bus}}-g_{\text{eff}}^{Q-\text{bus}}\), respectively. These frequencies are shown by circles in Fig.~3(b). We solved the inverse problem for \(f_Q^{'}\), \(f_C^{'}\), and \(f_{\text{bus}}^{'}\) in Hamiltonian \eqref{eq2}. After including the couplings \(g_{\text{eff}}^{Q-C}\), \(g_{\text{eff}}^{C-\text{bus}}\), and \(g_{\text{eff}}^{Q-\text{bus}}\), the model reproduces the experimentally measured dressed-state frequencies. The solution of this inverse problem is shown by the dashed line in Fig.~2. In the full-system method, \(D\) is obtained from a reverse projection. We calculate how much the non-computational dressed state \(|C\rangle\) contains the basis vectors \(|e,g,0\rangle\) and \(|g,g,1\rangle\):
$
D = \left| \langle e, g, 0 | C \rangle \right|^2
+ \left| \langle g, g, 1 | C \rangle \right|^2
= |c_{e, g, 0}|^2 + |c_{g, g, 1}|^2.
$
Because the interaction is symmetric, this is equivalent to estimating the contribution of \(|g,e,0\rangle\) to the dressed computational states \(|Q\rangle\) and \(|bus\rangle\). For the mode configuration in Fig.~4(b), the full-system method gives \(D=31\%\).

The second method uses an effective mode, as shown in Fig.~4(c). Instead of treating the resonant Q-bus pair explicitly, we replace it with a single effective bus mode. This mode is shifted by the effective interaction \(g_{\text{eff}}^{Q-\text{bus}}\) toward mode C. In the spectrum of the reduced original system \eqref{eq2}, this corresponds to the mode of the Q-bus avoided crossing that lies closest to C. The effective Hamiltonian is
\begin{equation}\label{eq3} 
\hat{H}_{\text{eff}}=h\begin{pmatrix}
f_C^{''} & g_{\text{eff}}^{C-\text{bus}}\\
g_{\text{eff}}^{C-\text{bus}} & f_{\text{bus}}^{''}
\end{pmatrix}.
\end{equation}

The dressed states \(|C\rangle\) and \(|bus\rangle\), which are the eigenvectors of Hamiltonian \eqref{eq3}, have frequencies \(f_C\) and \(f_{\text{bus}}+g_{\text{eff}}^{Q-\text{bus}}\). In this case, \(D\) quantifies the dressing of the single computational state \(|bus\rangle\) as $D = \left| \langle e, 0 | bus \rangle \right|^2=|c_{e, 0}|^2$.
For the configuration in Fig.~4(c), this method gives \(D=49\%\).

The third method uses an unperturbed mode, as shown in Fig.~4(d). It uses the same type of effective Hamiltonian \(\hat{H}_{\text{eff}}\) as the second method, but places the dressed computational state \(|bus\rangle\) at the center of the Q-bus avoided crossing of the reduced original system \eqref{eq2}. This effective Hamiltonian has its own unperturbed frequencies. Its eigenvalues are \(h\cdot f_C\) and \(h\cdot f_{\text{bus}}\), with eigenvectors \(|C\rangle\) and \(|bus\rangle\), respectively. Here \(D\) is estimated in the same way as in the second method, i.e. $D = \left| \langle e, 0 | bus \rangle \right|^2=|c_{e, 0}|^2$.
For the configuration in Fig.~4(d), the result is \(D=26\%\).

After illustrating the three methods at one value of \(f_U\), namely the point shown in Fig.~3(c), we compare them over the full frequency range in Fig.~4(a). The full-system method shows that the dressing \(D\) of the computational states \(|Q\rangle\) and \(|L\rangle\) by the non-computational state associated with \(|U\rangle\) increases from 1\% at \(f_U\approx4.4\) GHz to 50\% at \(f_U=f_U^c\approx4.252\) GHz. It then decreases back to 1\% at \(f_U\approx4.238\) GHz.

The full-system method also identifies the dominant basis vectors of \(|U\rangle\) and \(|L\rangle\) over the entire frequency range. For \(f_U>f_U^c\), the lower mode satisfies \(|L\rangle \leftrightarrow |bus\rangle \approx|g, g, 1\rangle\), as shown by the orange region in Fig.~4(a). For \(f_U<f_U^c\), the system crosses the \(D=50\%\) boundary and reaches \(|L\rangle \leftrightarrow |C\rangle \approx|g, e, 0\rangle\), shown by the green region. In the full-system method, crossing \(D=50\%\) occurs simultaneously with the inversion of the effective-coupling relation between \(g_{\text{eff}}^{Q-U}\) and \(g_{\text{eff}}^{Q-L}\) in Fig.~3(a). It also coincides with hybridization of mode L down to \(\omega_{\text{bus}}/2\pi-g_{C-\text{bus}}\), as shown in Fig.~2. We therefore identify the region near \(f_U=f_U^c\) as the region where the dominant basis vectors of \(|U\rangle\) and \(|L\rangle\) are exchanged. For \(f_U>f_U^c\), the correspondence is \(|U\rangle \leftrightarrow |C\rangle \approx|g, e, 0\rangle\) and \(|L\rangle \leftrightarrow |bus\rangle \approx|g, g, 1\rangle\). For \(f_U<f_U^c\), it becomes \(|U\rangle \leftrightarrow |bus\rangle \approx|g, g, 1\rangle\) and \(|L\rangle \leftrightarrow |C\rangle \approx|g, e, 0\rangle\).

Compared with the full-system method, the second and third methods slightly underestimate \(D\) in the high-frequency region. Near \(f_U=f_U^c\), the effective-mode method begins to overestimate \(D\). It crosses the 50\% threshold too early and leaves it too late. The unperturbed-mode method, by contrast, never reaches 50\%. Thus, the effective-mode method can be used as a conservative upper estimate of non-computational-state dressing near the critical point. This is useful when the goal is to remain in the dispersive regime.

All three methods can be used away from the critical point. The effective-mode and unperturbed-mode methods are especially useful in systems with many non-computational states, where they simplify the estimate of non-computational-state dressing. For a system with several couplers \(C_1,\ldots,C_N\), dressing of an effective or unperturbed computational state can be estimated as
$
\sum_{\mathcal{C}_1,...\mathcal{C}_N}
\left| \langle \mathcal{C}_1,...\mathcal{C}_N, 0 | bus \rangle \right|^2
=
\sum_{\mathcal{C}_1,...\mathcal{C}_N}
|c_{\mathcal{C}_1,...\mathcal{C}_N, 0}|^2.
$
The reverse projection used in the first method no longer works straightforwardly when several couplers are involved. This is a clear advantage of the second and third methods for reducing parasitic dressing of the computational subspace.

\section{Discussion and Conclusion}

The three-element system in Fig.~1 allowed us to study, in detail, how effective couplings, mode frequencies, and the bare-state composition of dressed states are related. This was possible because the coupling constants had a favorable ratio: the coupling between the scanning element Q and its neighboring element C was smaller than the coupling between the scanned elements C and bus, \(g_{Q-C}<g_{C-\text{bus}}\). As a result, we could resolve the positions of modes U and L and measure the effective Q-U and Q-L couplings up to the critical value \(D=50\%\) at \(f_U=f_U^c\). The scanning procedure did not significantly perturb the spectrum.

Figures~3(a) and 4(a) show that \(D\) strongly affects the effective interactions in the system. The local capacitive coupling \(g_{\text{eff}}^{Q-C}\) between Q and C decreases from the coupling constant \(g_{Q-C}\) at large detuning between the C and bus modes to the critical value \(g_{\text{eff}}^c\) at a detuning of \(2g_{C-\text{bus}}\), as shown by the sky-blue dashed line in Fig.~3(a). At the same time, the nonlocal coupling \(g_{\text{eff}}^{Q-\text{bus}}\) between Q and bus increases from the direct Q-bus coupling to \(g_{\text{eff}}^c\), as shown by the brown dashed line in Fig.~3(a). This behavior shows a direct connection between bare-state composition and effective coupling. The effective coupling between spatially local states, \(g_{\text{eff}}^{Q-C}\), dominates over the coupling between nonlocal states, \(g_{\text{eff}}^{Q-\text{bus}}\), throughout the full frequency range except at the critical point, where the two couplings become equal.

As discussed in Section~1, the inset of Fig.~2 shows hybridization of the bus mode from \(f_{\text{bus}}=\omega_{\text{bus}}/2\pi\) at low \(D\) to \(f_{\text{bus}}=\omega_{\text{bus}}/2\pi-g_{C-\text{bus}}\) at high \(D\). The bus-mode frequency changes within well-defined limits and correlates with increasing dressing by C. If the bus had an individual readout resonator, this frequency shift could be measured directly and used to characterize non-computational-state dressing of the target computational states.

The three methods for estimating \(D\) each have advantages and limitations. The first method is the most complete, but it is not directly applicable to systems with more than one non-computational state. The other two methods do not depend on the number of dressing states. The effective-mode method provides a good upper estimate of \(D\) and is fairly accurate at low \(D\). The unperturbed-mode method is reliable only at low \(D\), because it underestimates non-computational-state dressing near the critical value.

We now return to the analogy with universal quantum computing introduced in Section~2. Consider a two-qubit operation, such as iSWAP, between the computational states of Q and bus. The interaction is created by tuning coupler C, which represents the non-computational state \(|g,e,0 \rangle\) and approaches the computational states from above in frequency. According to the full-system estimate of \(D\), such an operation can use effective couplings \(g_{\text{eff}}^{Q-\text{bus}}\) up to \(g_{\text{eff}}^c\approx5.2\) MHz, shown by the brown line in Fig.~3(a). As the effective coupling increases, the participation of the non-computational state \(|g,e,0 \rangle\) in the dressed computational states \(|Q \rangle\) and \(|bus \rangle\) also increases, as shown by the orange region in Fig.~4(a).

If \(D\) is related to the coherence properties of the computational states through \(T_{1,\text{eff}}(D)\), in analogy with \(T_{1/\phi,\text{eff}}(p_{i,k})\) analysis attempt in Ref.~\cite{Marxer2023}, then the relaxation-induced error
$
\epsilon=\tau/T_{1,\text{eff}}(D)=\tau/T_{1,\text{eff}}(g_{\text{eff}}^{Q-\text{bus}})
$
no longer depends linearly on the operation duration \(\tau=1/4g_{\text{eff}}^{Q-\text{bus}}\) \cite{Zhong2021}. Therefore, when optimizing gate fidelity \cite{Yan2018}, one must account for dressing of the computational subspace by non-computational states.

Accurate estimation and control of \(D\) therefore directly affect gate fidelity in universal quantum computing. Comparing the first and third methods also gives another useful conclusion. As discussed in Section~2, at high effective coupling, \(D\) in the full system is larger than \(D\) obtained from the unperturbed-mode estimate for the same position of the coupler, as shown by the red dashed line in Fig.~4(a). This means that the coherence properties of computational states cannot be characterized accurately by measuring unperturbed modes separately at a fixed detuning from the non-computational mode. The characterization must be performed directly at the operating point of the two-qubit gate, with the corresponding spectral ordering, as in Refs.~\cite{Marxer2025,Marxer2023}.

\section*{Acknowledgements}

The authors acknowledge A. V. Ustinov for supporting this work and for valuable recommendations during the research. This work was supported by the Strategic Academic Leadership Program "Priority-2030" of the National University of Science and Technology MISIS (within the strategic technological project "Quantum Internet"). The numerical simulation of qubit interactions and spectroscopic measurements were funded by the Russian Science Foundation, Grant No. 26-12-00445. 


\end{document}